\documentclass[twocolumn,superscriptaddress,letter]{revtex4}%
\usepackage{amssymb}
\usepackage{color}
\usepackage{graphicx}
\usepackage{dcolumn}
\usepackage{bm}
\usepackage[header,title,page,titletoc]{appendix}
\usepackage{amsmath}
\usepackage{amsfonts}%
\providecommand{\U}[1]{\protect \rule{.1in}{.1in}}
\begin{document}
\title{Ferromagnetism and Antiferromagnetism of Correlated Topological Insulator with
Flat Band}
\author{Jing He}
\affiliation{Department of Physics, Beijing Normal University, Beijing 100875, China}
\author{Bo Wang}
\affiliation{Department of Physics, Beijing Normal University, Beijing 100875, China}
\author{Su-Peng Kou}
\thanks{Corresponding author}
\email{spkou@bnu.edu.cn}
\affiliation{Department of Physics, Beijing Normal University, Beijing 100875, China}

\begin{abstract}
In this paper, based on the mean field approach and
random-phase-approximation, we study the magnetic properties of the spinful
Haldane model on honeycomb lattice of topological flat band with on-site
repulsive Coulomb interaction. We find that the antiferromagnetic (AF) order
is more stable than ferromagnetic (FM) order at (or near) half-filling; while
away from half-filling the phase diagram becomes complex: At large doping, FM
order is more stable than AF order due to the flatness of band structure. In
particular, we find that at quarter filling case, the system becomes a $Q=1$
topological insulator which is induced by the FM order.

\end{abstract}
\maketitle

\section{Introduction}

From 1980 year, topological ordered states become a hot topic and a lot of
efforts had been done on this issue\cite{wen}. Integer and fractional quantum
Hall effect (IQHE and FQHE) have attracted great attention and been studied a
great deal since first discovered in two-dimensional electron gas with Laudau
levels induced by strong magnetic field\cite{2,t}. Haldane's honeycomb
model\cite{Haldane} demonstrates IQHE in lattice model without Landau levels
while FQHE had not been discovered in lattice models until recently\cite{tang,
sun, neupert}. The topologically nontrivial flat-land was pointed to be key in
realizing FQHE in lattice models based on the mechanism of quadratic band
touching. And the lattice models are designed reaching a high flatness ratio
(the ratio of the band gap over bandwidth) about $50$ in different lattice
models (kag\'{o}me lattice, square lattice and honeycomb lattice)\cite{tang,
sun, neupert}. In these models, the hoppings of next-nearest neighbor (NNN) or
next-next-nearest neighbor (NNNN) besides nearest neighbor (NN) are introduced
to achieve the flat-bands. And the flat-bands of these models are shown to be
with non-trivial topology (with non-zero Chern number). To obtain FQHE states,
people always consider these tight-binding models of spinless fermions or
bosons with the NN and the NNN repulsions under a hard-core condition. Then
the numerical results confirm the existence of topologically nontrivial
flat-band\cite{sheng,WYF}.

On the other hand, the spin orders are important magnetic properties in a
strongly correlated electron system, on which people had also paid much
attention\cite{Auerbach}. For example, an interesting problem is about the
ferromagnetic (FM) order in correlated electron system with flat band.
According to the Stoner criteria, for correlated electronic system with flat
or almost flat band, the Coulomb interaction will remove the band degeneracy
and drive the system into a ferromagnetic ground state\cite{mi,ta}. So we may
ask a question : in a spinful fermionic lattice model of topologically
nontrivial flat-band with on-site Coulomb interaction, whether the ground
state is FM order? This is the starting point of this paper.

After doing mean field calculations, in a typical two-component fermionic
lattice model of topologically nontrivial flat-band with on-site Coulomb
interaction (the spinful Haldane model on honeycomb
lattice\cite{Haldane,he1,he2}), we find that the AF order is more stable than
FM order at (or near) half-filling; while away from half-filling the phase
diagram becomes complex: At large doping, FM order is more stable than AF
order due to the flat band. In particular, we found that at quarter filling
case, the system becomes a $Q=1$ topological insulator which is induced by the
FM order. In Ref.\cite{ka}, it was proved that the ground state of the Hubbard
model with topological flat-bands is a ferromagnetic order when the lowest
flat band is half-filled. Our results are consistent with this rigorous
provement. In addition, in Ref.\cite{shen}, quantum anomalous Hall effect in a
flat-band ferromagnet on a two dimensional decorated lattice with spin-orbit
coupling was also predicted.

The paper is organized as follows. Firstly, we write down the Hamiltonian of
the spinful Haldane model on honeycomb lattice with topological flat band in
Sec.II. In Sec.III, we do mean field calculations and get the phase diagram.
In Sec.IV, we discuss magnetic properties and topological properties for the
the half filling case. Then in Sec.V, we discuss the magnetic properties and
the topological properties for the the quarter filling case. Next we discuss
the other doping cases in VI. Finally, the conclusions are given in Sec.VII.

\begin{figure}[ptb]
\includegraphics[width=0.55\textwidth]{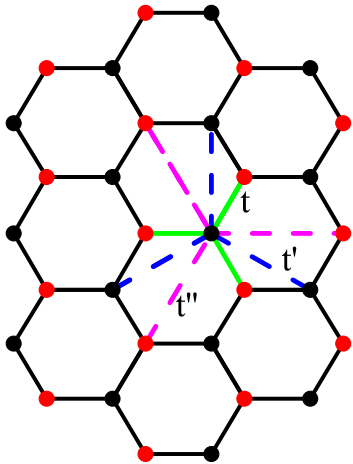}\caption{(Color online) The
illustration of the honeycomb lattice. The red and black dots represent the
bipartite lattice. The green lines, dash blue lines and dash magenta lines are
the nearest hopping, $t$, the next nearest hopping $t^{\prime}$ and the next
next nearest hopping, $t^{\prime \prime}$, respectively.}%
\end{figure}

\section{The model Hamiltonian}

In this paper we start from the spinful Haldane model with the on-site Coulomb
interaction with the Hamiltonian as\cite{Haldane,he1,he2}
\begin{align}
H  &  =-t\sum \limits_{\left \langle {i,j}\right \rangle ,\sigma}\hat{c}%
_{i\sigma}^{\dagger}\hat{c}_{j\sigma}-t^{\prime}\sum \limits_{\left \langle
\left \langle {i,j}\right \rangle \right \rangle ,\sigma}e^{i\phi_{ij}}\hat
{c}_{i\sigma}^{\dagger}\hat{c}_{j\sigma}\nonumber \\
&  -t^{\prime \prime}\sum \limits_{\left \langle \left \langle \left \langle
{i,j}\right \rangle \right \rangle \right \rangle ,\sigma}\hat{c}_{i\sigma
}^{\dagger}\hat{c}_{j\sigma}+U\sum \limits_{i}\hat{n}_{i\uparrow}\hat
{n}_{i\downarrow}\nonumber \\
&  -\mu \sum \limits_{i,\sigma}\hat{c}_{i\sigma}^{\dagger}\hat{c}_{i\sigma}+h.c.
\end{align}
where $U$ is the on-site Coulomb repulsive interaction strength and $\mu$ is
the chemical potential. $t$, $t^{\prime}$ and $t^{\prime \prime}$ are the
nearest neighbor, the next nearest neighbor and the next next nearest neighbor
hoppings, respectively (See Fig.1). We all know that the honeycomb lattice is
a bipartite lattice which is represent by the red and black dots in Fig.1, and
the green lines, dash blue lines and dash magenta lines are the nearest
hoppings, the next nearest hoppings and the next next nearest hoppings,
respectively. In the flat-band limit we have $t^{\prime}=0.6t,$ $t^{\prime
\prime}=-0.58t$\cite{neupert}. $\phi_{ij}=0.4\pi$ is a complex phase to the
next nearest neighbor hopping, of which the positive phase is set to be
clockwise\cite{neupert}.

\begin{figure}[ptb]
\includegraphics[width=0.55\textwidth]{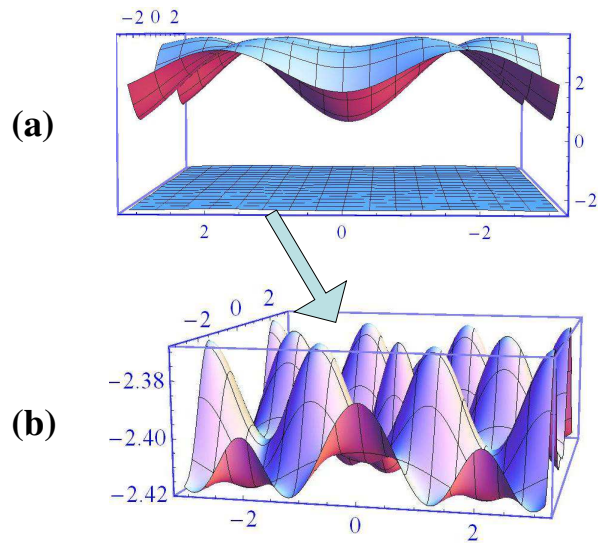}\caption{(Color online) The
free electronic dispersion ($U/t=0$) at $t^{\prime}=0.6t,$ $t^{\prime \prime
}=-0.58t$ and $\phi_{ij}=0.4\pi$. (a) is the two bands, of which each band is
doubly degenerate. (b) shows the details of the flat band. One can see that
the bandwidth is much smaller than the band gap. }%
\end{figure}

\begin{figure}[ptb]
\includegraphics[width=0.5\textwidth]{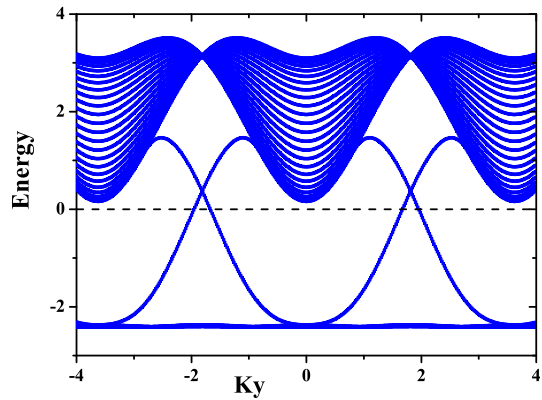}\caption{(Color online) The
Zigzag edge states along x-direction at $U/t=0$, $t^{\prime}=0.6t,$
$t^{\prime \prime}=-0.58t$ and $\phi_{ij}=0.4\pi.$ The black dash line is the
position of chemical potential. }%
\end{figure}

For the free electrons case ($U/t=0$), the lower energy band becomes flat with
nonzero Chern numbers $Q=2$ and the flatness ratio is about $50$ (See Fig.2).
From Fig.2, we can see that (a) is the totally energy bands, spin up and spin
down are degenerate due to the spin-rotation symmetry and (b) is the
amplification of the flat band. The flat band still has dispersion but the
bandwidth is much smaller than the band gap. Fig.3 illustrates the existence
of edge states in this $Q=2$ topological insulator.

\section{Mean field approach}

Next, we consider the interacting case of $U/t\neq0$. An interesting question
is : \emph{what is the ground state for the spinful Haldane model with the
on-site interaction in the TFB limit when the }$U/t\neq0?$ As discussed above,
people know that for the case of an interacting fermion model with flat-band,
the ground state is always a FM order. On the other hand, near the half
filling, the super-exchange effect will lead to an antiferromagnetic (AF) spin
interaction between two local spin moments nearby. Thus to answer this
question, we will study the competition between AF order and FM order by the
following mean-field ansatzs :
\begin{equation}
\langle \hat{c}_{i,\sigma}^{\dagger}\hat{c}_{i,\sigma}\rangle=\frac{1}%
{2}\left[  n+(-1)^{i}\sigma M_{AF}\right]
\end{equation}
and
\begin{equation}
\langle \hat{c}_{i,\sigma}^{\dagger}\hat{c}_{i,\sigma}\rangle=\frac{1}%
{2}(n+\sigma M_{F})
\end{equation}
where $M_{AF}$ is the staggered antiferromagnetic order parameter and $M_{F}$
is the ferromagnetic order parameter, $n$ is the average particle number on
each site. For the case of spin-up and spin-down, we have $\sigma=1$ and
$\sigma=-1$, respectively.

\subsection{AF order}

Firstly we consider the instability of AF order. For the AF order, we have
$M_{AF}\neq0,$ now the effective Hamiltonian can be written as:
\begin{align}
H  &  =-t\sum \limits_{\left \langle {i,j}\right \rangle ,\sigma}\hat{c}%
_{i\sigma}^{\dagger}\hat{c}_{j\sigma}-t^{\prime}\sum \limits_{\left \langle
\left \langle {i,j}\right \rangle \right \rangle ,\sigma}e^{i\phi_{ij}}\hat
{c}_{i\sigma}^{\dagger}\hat{c}_{j\sigma}-t^{\prime \prime}\sum
\limits_{\left \langle \left \langle \left \langle {i,j}\right \rangle
\right \rangle \right \rangle ,\sigma}\hat{c}_{i\sigma}^{\dagger}\hat
{c}_{j\sigma}\nonumber \\
&  -\sum_{i}\left(  -1\right)  ^{i}\Delta_{A}\hat{c}_{i}^{\dagger}\sigma
_{z}\hat{c}_{i}+\sum_{i,\sigma}\left(  \frac{Un}{2}-\mu \right)  \hat
{c}_{i\sigma}^{\dagger}\hat{c}_{i\sigma}+h.c.
\end{align}
where $\Delta_{A}=UM_{AF}/2$ and $\sigma_{z}$\ is Pauli matrix. $n=1-d$ is the
average particle number on each site and $d$ is the hole concentration. After
Fourier transformation, the Hamiltonian in momentum space becomes
\begin{equation}
H=\sum \limits_{k}\Psi_{k}^{\dagger}h_{Ak}\Psi_{k}%
\end{equation}
where $\Psi_{k}^{\dagger}=\left(
\begin{array}
[c]{cccc}%
\hat{a}_{k\uparrow}^{\dagger} & \hat{b}_{k\uparrow}^{\dagger} & \hat
{a}_{k\downarrow}^{\dagger} & \hat{b}_{k\downarrow}^{\dagger}%
\end{array}
\right)  $ and \begin{widetext}
\[
h_{Ak}=\left(
\begin{array}
[c]{cccc}%
-t^{\prime}\left(  C-D\right)  +\Delta_{A} & \xi_{k} & 0 & 0\\
\xi_{k}^{\ast} & -t^{\prime}\left(  C+D\right)  -\Delta_{A} & 0 & 0\\
0 & 0 & -t^{\prime}\left(  C-D\right)  -\Delta_{A} & \xi_{k}\\
0 & 0 & \xi_{k}^{\ast} & -t^{\prime}\left(  C+D\right)  +\Delta_{A}%
\end{array}
\right)
\]
\end{widetext}where $\xi_{k}=-t\left(  A+iB\right)  -t^{\prime \prime}\left(
E+iF\right)  $.

Then we can get the spectrum of the electrons as
\begin{equation}
E_{A,k,1,\pm}=-t^{\prime}C+\sqrt{\left \vert \xi_{k}\right \vert ^{2}+\left(
t^{\prime}D\pm \Delta_{A}\right)  ^{2}}%
\end{equation}%
\begin{equation}
E_{A,k,2,\pm}=-t^{\prime}C-\sqrt{\left \vert \xi_{k}\right \vert ^{2}+\left(
t^{\prime}D\pm \Delta_{A}\right)  ^{2}}%
\end{equation}
where
\begin{align*}
A  &  =2\cos \left(  k_{x}/2\right)  \cos \left(  \sqrt{3}k_{y}/2\right)  +\cos
k_{x}\\
B  &  =2\sin \left(  k_{x}/2\right)  \cos \left(  \sqrt{3}k_{y}/2\right)  -\sin
k_{x}\\
C  &  =2\cos \phi \sum_{i=1}^{3}\cos \left(  \mathbf{k\cdot b}_{i}\right) \\
D  &  =2\sin \phi \sum_{i=1}^{3}\sin \left(  \mathbf{k\cdot b}_{i}\right) \\
E  &  =2\cos k_{x}\cos \left(  \sqrt{3}k_{y}\right)  +\cos2k_{x}\\
F  &  =-2\sin k_{x}\cos \left(  \sqrt{3}k_{y}\right)  +\sin2k_{x}%
\end{align*}
where $\mathbf{b}_{i}$ is the next nearest neighbor vectors.

At different doping, by minimizing the free energy of the mean field, the
self-consistent equations at zero temperature in the reduced BZ
are\cite{Peres,kou4}:%
\begin{align}
1  &  =\frac{1}{N_{s}M_{AF}}[\sum_{E_{A,k,2,+}<\mu_{_{eff}}}\frac{t^{\prime
}D+\Delta_{A}}{2\xi_{k1}}\nonumber \\
&  -\sum_{E_{A,k,2,-}<\mu_{_{eff}}}\frac{t^{\prime}D-\Delta_{A}}{2\xi_{k2}%
}],\\
1-d  &  =\frac{1}{2N_{s}}[\sum_{E_{A,k,2,+}<\mu_{_{eff}}}1+\sum_{E_{A,k,2,-}%
<\mu_{_{eff}}}1]
\end{align}
where $\mu_{_{eff}}=\mu-\frac{Un}{2}$ and $N_{s}$ is the number of the unit
cells and%
\[
\xi_{k1}=\sqrt{\left \vert \xi_{k}\right \vert ^{2}+\left(  t^{\prime}%
D+\Delta_{A}\right)  ^{2}},
\]%
\[
\xi_{k2}=\sqrt{\left \vert \xi_{k}\right \vert ^{2}+\left(  t^{\prime}%
D-\Delta_{A}\right)  ^{2}}.
\]

In Fig.4, we give the electronic dispersions for AF order at $U/t=7.0$ for
half-filling. Comparing with Fig.2 and Fig.4, we can see that AF order
parameter can change the flat band into a dispersive one.

\begin{figure}[ptb]
\includegraphics[width=0.5\textwidth]{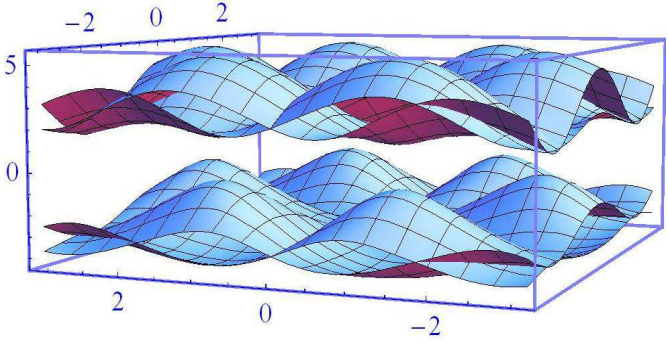}\caption{(Color online) The
electronic dispersions for AF order at $U/t=7.0,$ $t^{\prime}=0.6t,$
$t^{\prime \prime}=-0.58t$ and $\phi_{ij}=0.4\pi$ for half filling. We can see
that the AF order parameter changes the flat band into a dispersive one.}%
\end{figure}

\subsection{FM order}

Secondly we consider the instability of FM order. For FM order, we have
$M_{F}\neq0,$ now the effective Hamiltonian can be written as
\begin{align}
H  &  =-t\sum \limits_{\left \langle {i,j}\right \rangle ,\sigma}\hat{c}%
_{i\sigma}^{\dagger}\hat{c}_{j\sigma}-t^{\prime}\sum \limits_{\left \langle
\left \langle {i,j}\right \rangle \right \rangle ,\sigma}e^{i\phi_{ij}}\hat
{c}_{i\sigma}^{\dagger}\hat{c}_{j\sigma}-t^{\prime \prime}\sum
\limits_{\left \langle \left \langle \left \langle {i,j}\right \rangle
\right \rangle \right \rangle ,\sigma}\hat{c}_{i\sigma}^{\dagger}\hat
{c}_{j\sigma}\nonumber \\
&  -\sum_{i}\Delta_{F}\hat{c}_{i}^{\dagger}\sigma_{z}\hat{c}_{i}%
+\sum_{i,\sigma}\left(  \frac{Un}{2}-\mu \right)  \hat{c}_{i\sigma}^{\dagger
}\hat{c}_{i\sigma}+h.c.
\end{align}
where $\Delta_{F}=$ $UM_{F}/2.$ After Fourier transformation, the Hamiltonian
in momentum space becomes
\begin{equation}
H=\sum \limits_{k}\Psi_{k}^{\dagger}h_{Fk}\Psi_{k}%
\end{equation}
where \begin{widetext}
\[
h_{Fk}=\left(
\begin{array}
[c]{cccc}%
-t^{\prime}\left(  C-D\right)  -\Delta_{F} & \xi_{k}  & 0 & 0\\
\xi_{k}^{\ast}  &
-t^{\prime
}\left(  C+D\right)  -\Delta_{F} & 0 & 0\\
0 & 0 & -t^{\prime}\left(  C-D\right)  +\Delta_{F} & \xi_{k} \\
0 & 0 &\xi_{k}^{\ast}
&
-t^{\prime}\left(  C+D\right)  +\Delta_{F}%
\end{array}
\right)
\]
\end{widetext}

Then we can get the spectrum of the electrons in FM order as
\begin{equation}
E_{F,k,1,\pm}=-t^{\prime}C-\Delta_{F}\pm \sqrt{\left \vert \xi_{k}\right \vert
^{2}+\left(  t^{\prime}D\right)  ^{2}}%
\end{equation}%
\begin{equation}
E_{F,k,2,\pm}=-t^{\prime}C+\Delta_{F}\pm \sqrt{\left \vert \xi_{k}\right \vert
^{2}+\left(  t^{\prime}D\right)  ^{2}}%
\end{equation}

At different doping, by minimizing the free energy of the mean field, the
self-consistent equations at zero temperature in the reduced BZ are%
\begin{align}
1  &  =\frac{1}{N_{s}M_{F}}[\sum_{E_{F,k,1,-}<\mu_{_{eff}}}\frac{1}%
{2}\nonumber \\
&  +\sum_{E_{F,k,1,+}<\mu_{_{eff}}}\frac{1}{2}-\sum_{E_{F,k,2,-}<\mu_{_{eff}}%
}\frac{1}{2}],\\
1-d  &  =\frac{1}{2N_{s}}[\sum_{E_{F,k,1,-}<\mu_{_{eff}}}1\nonumber \\
&  +\sum_{E_{F,k,1,+}<\mu_{_{eff}}}1+\sum_{E_{F,k,2,-}<\mu_{_{eff}}}1]
\end{align}
where $\mu_{_{eff}}=\mu-\frac{Un}{2}$ and $N_{s}$ is the number of the unit cells.

In Fig.5, we give the electronic dispersions for FM order at $U/t=7.0$ for
half-filling. Comparing with Fig.2 and Fig.5, we can see that FM order
parameter only splits the energy of the flat band without changing the
flatness ratio.

\begin{figure}[ptb]
\includegraphics[width=0.5\textwidth]{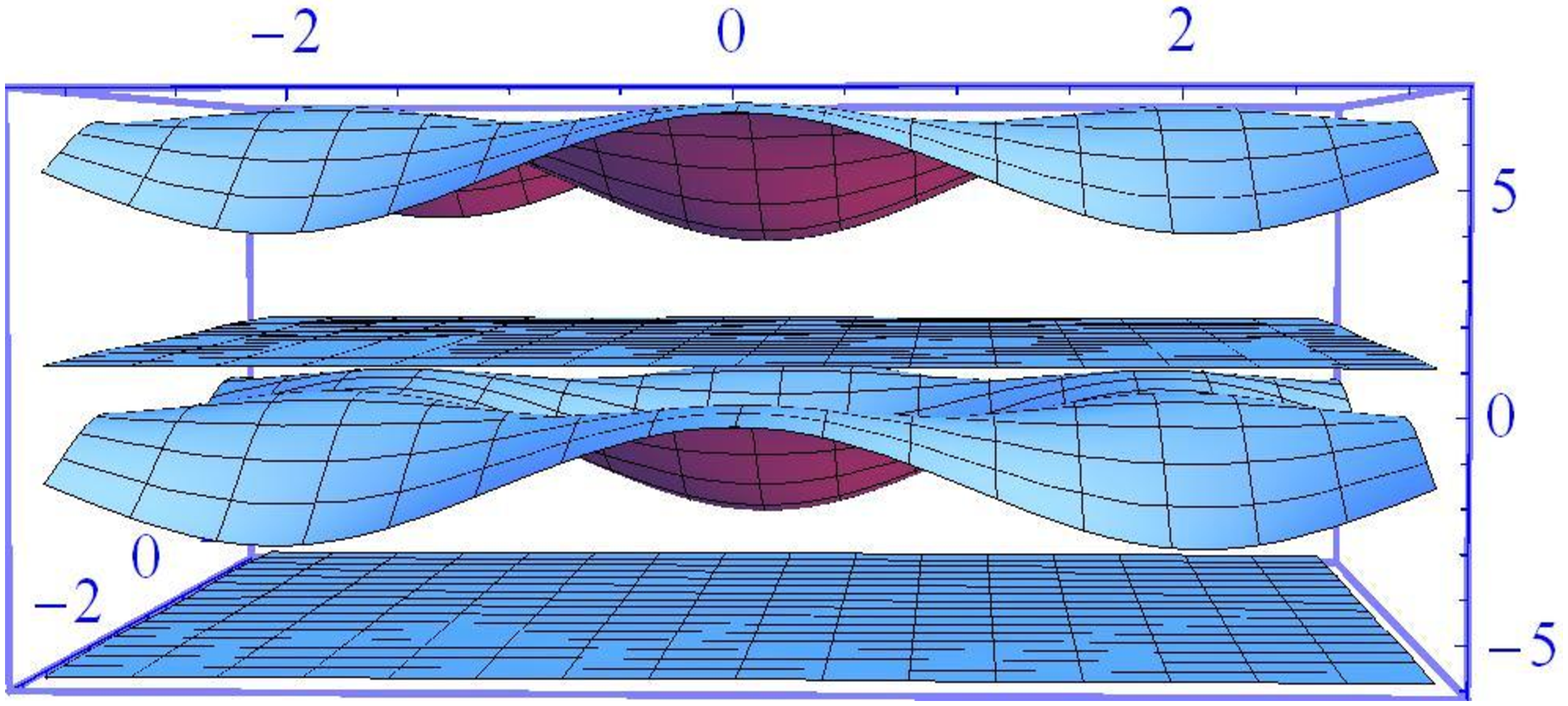}\caption{(Color online) The
electronic dispersions for FM order at $U/t=7.0,$ $t^{\prime}=0.6t,$
$t^{\prime \prime}=-0.58t$ and $\phi_{ij}=0.4\pi$ for half filling. We can see
that FM order parameter splits the two degenerate flat bands without changing
their flat ratio.}%
\end{figure}

\subsection{Global phase diagram}

\begin{figure}[ptb]
\includegraphics[width=0.5\textwidth]{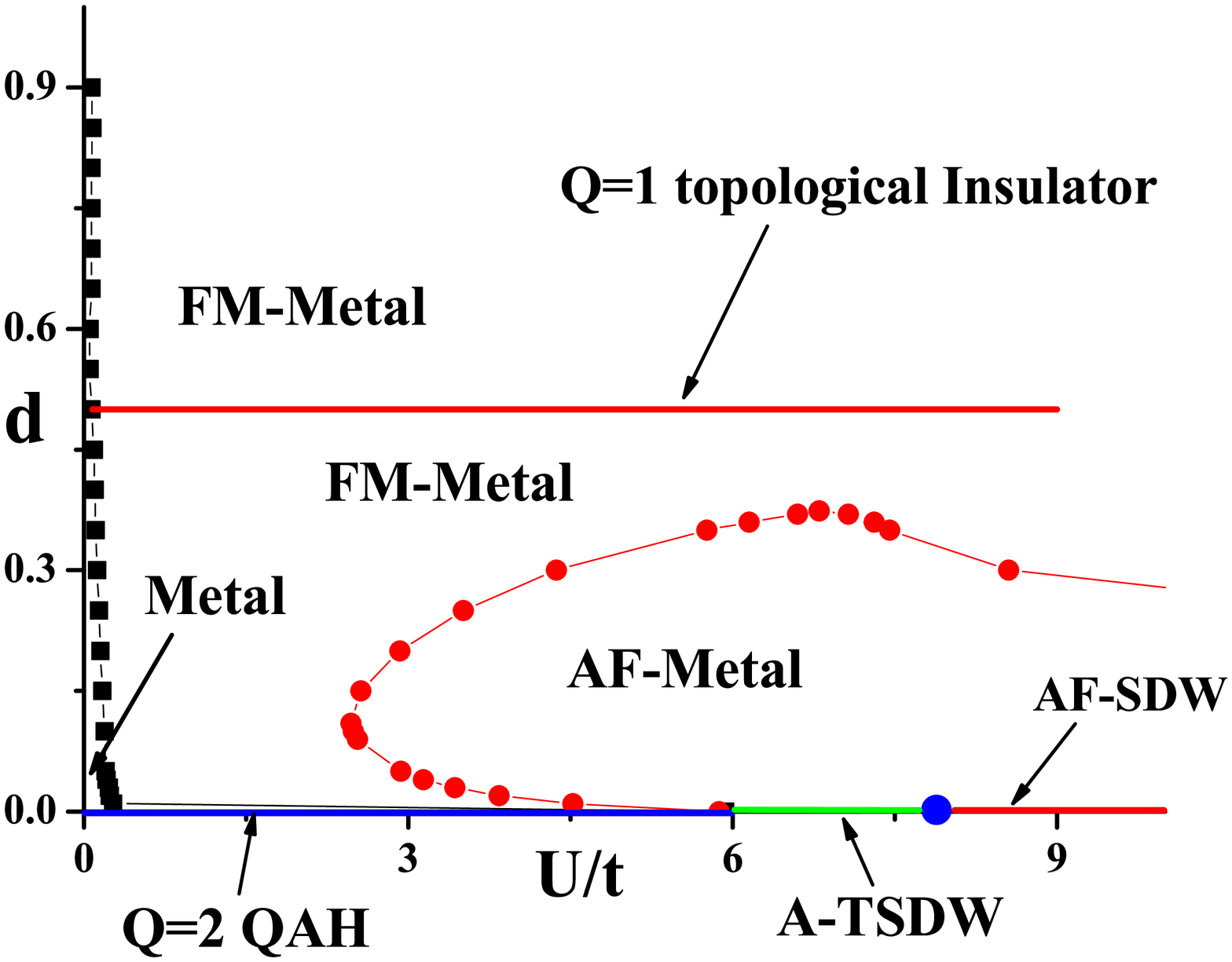}\caption{(Color online) The
global phase diagram for the parameters to be $t^{\prime}=0.6t,$
$t^{\prime \prime}=-0.58t$ and $\phi_{ij}=0.4\pi.$ At half filling case $d=0,$
there are three phases: $Q=2$ topological insulator with quantized anomalous
Hall effect, A-TSDW and trivial AF order. For the quarter filling case:
$d=0.5,$ there are two phases : paramagnetic Metal, $Q=1$ topological
Insulator with FM order. For other cases, there are three phases for the
region of $0<d<0.5$: paramagnetic Metal, FM-metal and AF-metal while there are
two phases for the region of $0.5<d<1$: paramagnetic Metal and FM-metal.}%
\end{figure}

Then by using mean field approach above we can get the global phase diagram
for different doping in Fig.6 when the parameters are $t^{\prime}=0.6t,$
$t^{\prime \prime}=-0.58t$, $\phi_{ij}=0.4\pi$. From Fig.6, we can see that
there exist seven phases: normal metal, FM-metal, AF-metal, $Q=1$ FM ordered
topological Insulator, $Q=2$ topological insulator with quantized anomalous
Hall effect (we may also call it QAH\ state),$\ $A-type antiferromagnetic-spin
density wave (A-type AF-SDW) topological insulator and trivial AF order.

At half filling case $d=0,$ in the weakly coupling limit $\left(
U/t<5.932\right)  $, the ground state is a $Q=2$ topological insulator with
flat-band. In the strong coupling region $\left(  U/t>5.932\right)  $, due to
$M_{AF}\neq0$, the ground state becomes an AF-SDW insulator. And at half
filling case, the AF order always has lower energy than FM order. When after
doping, the system is away from half filling, $d\neq0$. Now the repulsive
interaction is highly relevant for a fermion system with flat-band. And small
interaction will lead to an FM order. Thus in the phase diagram, away from
half filling ($d\neq0$), there exists a quantum phase transition between
paramagnetic metallic phase and\ FM order for little interaction strength due
to a big flatness ratio. And we find a narrow paramagnetic metallic phase. At
hole concentration $d=0.5,$ the situation becomes much different : the FM
order is really a $Q=1$ topological insulator. For other hole concentration
cases $d\neq0.5,$ $d\neq1,$ the ground state becomes a FM metal. When the
interaction strength becomes bigger, FM order is stable in a large region in
the global phase diagram due to the flatness of the fermion dispersion. Only
near half filling, the ground state energy for every site of AF order may be
lower than FM order at some intermediate interaction strength (See below). And
there exist first order quantum phase transitions between FM order and
AF\ order (See the red dot line in Fig.6).

In the following, we will give a detailed discussion on the cases of $d=0,$
$d=0.5$ and $d\neq0,$ $0.5,$ respectively.

\section{Half filling case : $d=0$}

Firstly we consider the half filling case. At half filling $d=0,$ the particle
number on average site is $n=1$ and the chemical potential $\mu=U/2$ $\left(
\mu_{_{eff}}=0\right)  .$ From Fig.6, we can see that at $d=0$ the ground
state is always dominated by the AF-SDW, for the ground state energy of AF
order per site is lower than the FM order (See Fig.7). In the weakly
interaction limit $\left(  U/t<5.932\right)  $, the ground state is a $Q=2$
topological insulator with flat-band. In the strong interaction region
$\left(  U/t>5.932\right)  $, due to $M_{AF}\neq0$, the ground state becomes
an AF-SDW insulator.

\begin{figure}[ptb]
\includegraphics[width=0.5\textwidth]{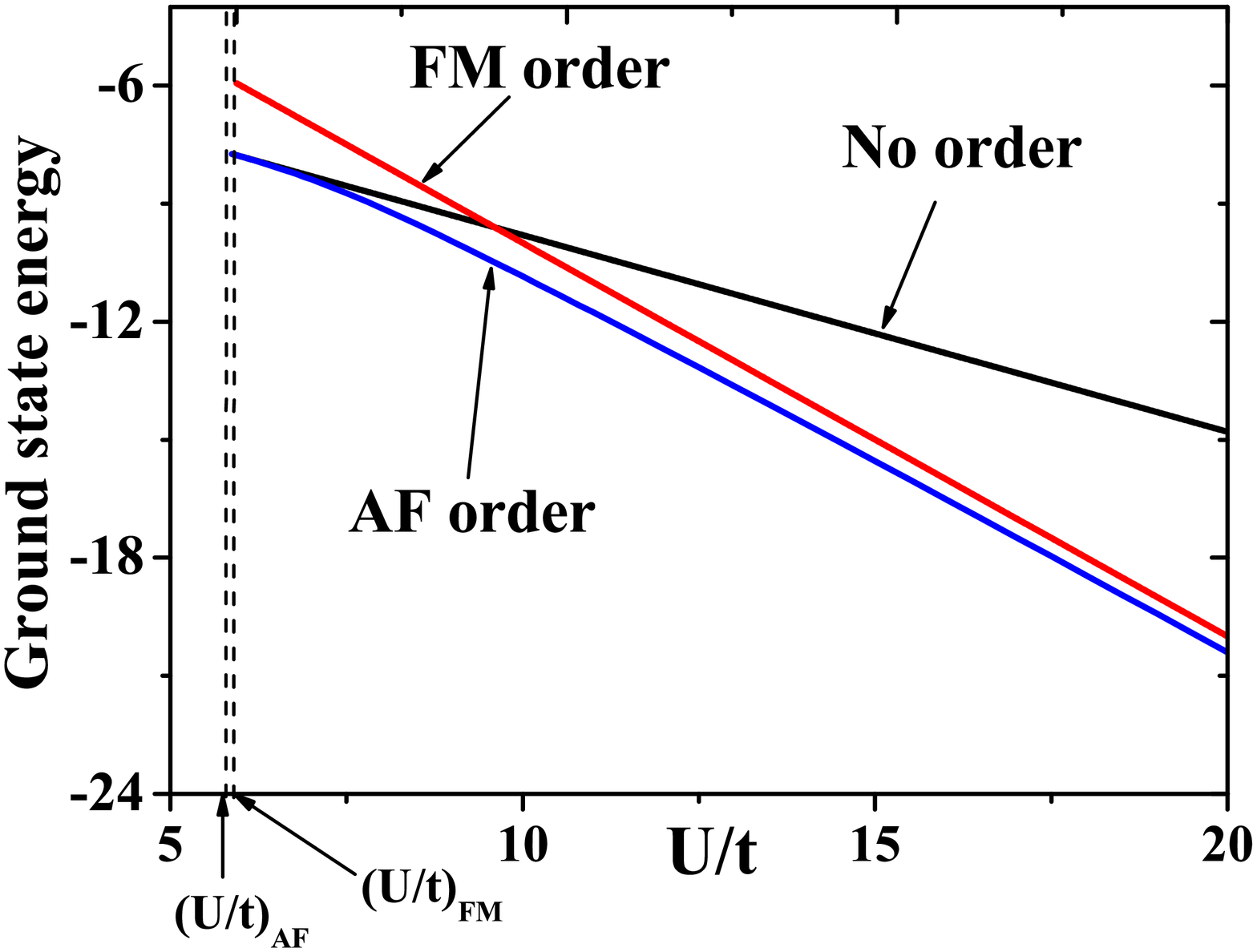}\caption{(Color online) The
ground state energies of every site for AF order (Green line), FM order (Red
line) and on spin order (black) per site at half filling for the parameters as
$t^{\prime}=0.6t,$ $t^{\prime \prime}=-0.58t$ $\phi_{ij}=0.4\pi.$}%
\end{figure}

\begin{figure}[ptb]
\includegraphics[width=0.55\textwidth]{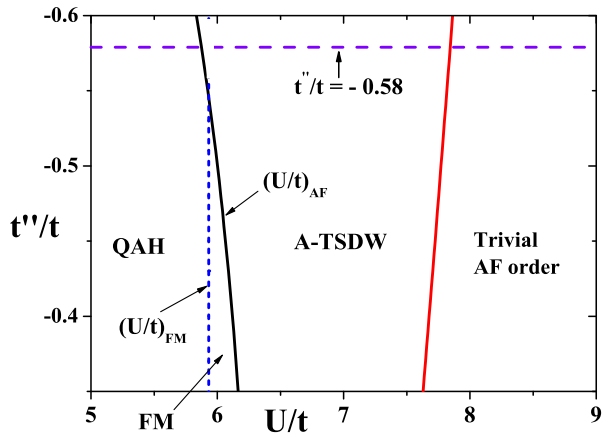}\caption{(Color online) The
phase diagram at half filling: there are four phases, QAH, FM-SDW, A-TSDW and
Trivial AF order. The dash blue line represent the FM phase transition
$(U/t)_{FM}$, the black line represent the AF phase transition $(U/t)_{AF},$
the red line represent a topological quantum phase transition. The purple dash
line is the case with largest flat ratio, $t^{\prime}=0.6t,$ $t^{\prime \prime
}=-0.58t$, $\phi_{ij}=0.4\pi$.}%
\end{figure}

Next we plot a phase diagram at half filling with different values of
$t^{\prime \prime}/t$ in Fig.8 (the purple dash line denotes $t^{\prime \prime
}=-0.58t,$ the dispersion of largest flat ratio). In Fig.8, there are three
quantum phase transitions $(U/t)_{FM},$ $(U/t)_{AF},$ and a topological
quantum phase transition (the red line) that divide four phases: $Q=2$
topological insulator, FM order, A-type topological AF-SDW (A-TSDW) and
trivial AF order. For the flat-band case (the purple dash line of
$t^{\prime \prime}=-0.58t$), one can see that the $Q=2$ topological insulator
occurs in the weak interaction, and with the increasing of $U/t,$ firstly the
system turns into A-TSDW state with $M_{AF}\neq0,$ and then turns into the AF
insulator with trivial topological properties. Now there only exist two types
of phase transitions: one is the magnetic phase transition (the black line in
Fig.8) between a magnetic order state with $M_{AF}\neq0$ and a non-magnetic
state with $M_{AF}=0$, the other is a topological quantum phase transition
(the red line in Fig.8) which is characterized by the condition of zero
fermion energy gaps (see the blue dot in Fig.6). And in Fig.9 and Fig.10, we
plot the staggered magnetization and the energy gap for the case with the
parameters $t^{\prime}=0.6t,$ $t^{\prime \prime}=-0.58t$ and $\phi_{ij}=0.4\pi$
at half filling.

\begin{figure}[ptb]
\includegraphics[width=0.5\textwidth]{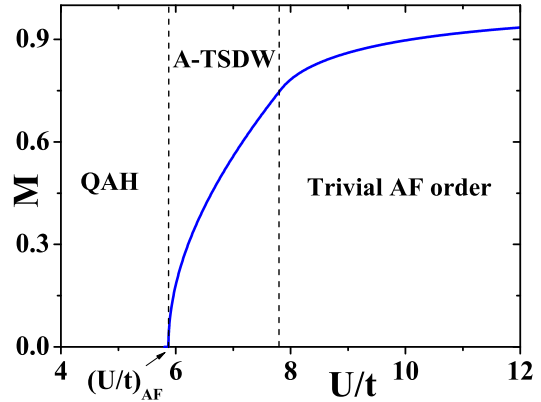}\caption{(Color online) The
staggered magnetization $M_{AF}$ for the half filling case with the parameters
$t^{\prime}=0.6t,$ $t^{\prime \prime}=-0.58t$ and $\phi_{ij}=0.4\pi.$ There are
two quantum phase transitions (the black dash lines) : one is the magnetic
phase transition at $(U/t)_{AF}$, the other is a topological quantum phase
transition with gap closing.}%
\end{figure}

\begin{figure}[ptb]
\includegraphics[width=0.5\textwidth]{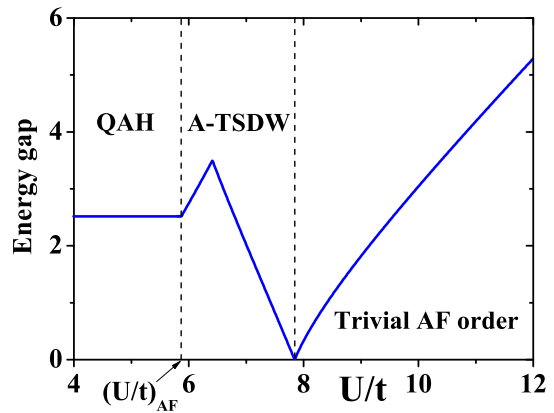}\caption{(Color online) The
energy gap of electrons for the half filling case with the parameters to be
$t^{\prime}=0.6t,$ $t^{\prime \prime}=-0.58t$ and $\phi_{ij}=0.4\pi.$ There are
two quantum phase transitions (the black dash lines) : one is the magnetic
phase transition at $(U/t)_{AF}$, the other is a topological quantum phase
transition with gap closing.}%
\end{figure}

In Fig.8, there is a dash blue line which is the boundary between $Q=2$
topological insulator and FM order. At half filling case, the ground state
energy of AF order per site is lower than the FM order. While away from
$t^{\prime \prime}=-0.58t$ (the dispersion of largest flat ratio), there may
still exist a narrow window of an FM order for $t^{\prime \prime}/t>-0.55.$

\begin{figure}[ptb]
\includegraphics[width=0.5\textwidth]{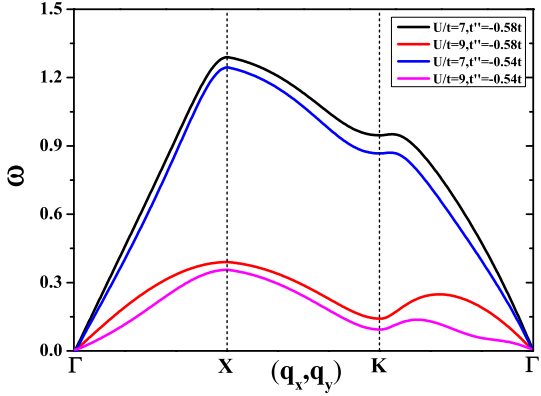}\caption{(Color online) The
dispersions of the spin wave of the AF order for the case of largest flat
ratio of the parameters are $t^{\prime}=0.6t$, $t^{\prime \prime}=-0.58t$,
$\phi_{ij}=0.4\pi$ with $U/t=7.0$ (balck line) and $U/t=9.0$ (red line) at
half filling. The dispersions of the spin wave of the AF order away from the
flat-band case are also given : $t^{\prime}=0.6t$, $t^{\prime \prime}=-0.54t$,
$\phi_{ij}=0.4\pi$ with $U/t=7.0$ (blue line) and $U/t=9.0$ (pink line). }%
\end{figure}

At the half filling case, when the AF order parameter is non-zero, the
electronic band structure will be reorganized. Thus in AF order, the spectrums
of electrons are dispersive. In addition, we go beyond the mean field theory
to study the collective spin fluctuations by random phase approximation (RPA)
method. We obtain the dispersion of spin excitations for the ground state. See
Fig.11. From Fig.11, one can see that in the AF order where $U/t>5.932,$
($U/t=9.0$ in trivial AF-SDW order and $U/t=7.0$ in A-TSDW order) there exist
gapless Goldstone modes - spin waves. We also calculate the dispersions of the
spin wave of the AF order away from the flat-band case, $t^{\prime}=0.6t$,
$t^{\prime \prime}=-0.54t$, $\phi_{ij}=0.4\pi$ with $U/t=7.0$ (blue line) and
$U/t=9.0$ (pink line). From this results, we can say that the spin waves are
dispersive and have linear behavior, $\omega(k)\propto q$ around $\Gamma$
pont, and also don't show anomalous behaviors at the flat band case
($t^{\prime \prime}=-0.58t$) in both trivial AF-SDW order and A-TSDW order.

\section{Quarter filling case : $d=0.5$}

\begin{figure}[ptb]
\includegraphics[width=0.5\textwidth]{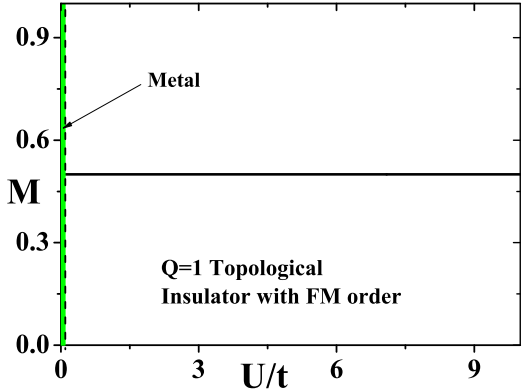}\caption{(Color online) The
staggered magnetization $M_{F}$ at the quarter filling ($d=0.5$) for the case
with the parameters $t^{\prime}=0.6t,$ $t^{\prime \prime}=-0.58t$ and
$\phi_{ij}=0.4\pi$. There have two phases: the normal paramagnetic metal (the
Green region) and the $Q=1$ topological insulator with FM order. In FM order,
the ferromagnetization saturates suddenly. }%
\end{figure}

\begin{figure}[ptb]
\includegraphics[width=0.5\textwidth]{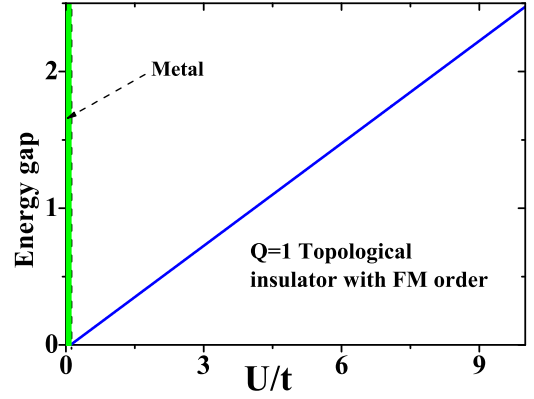}\caption{(Color online) The
electronic enegy gap at the quarter filling ($d=0.5$) for the case with the
parameters $t^{\prime}=0.6t,$ $t^{\prime \prime}=-0.58t$ and $\phi_{ij}=0.4\pi
$. There have two phases: the normal paramagnetic metal (the Green region) and
the $Q=1$ topological insulator with FM order.}%
\end{figure}

\begin{figure}[ptb]
\includegraphics[width=0.5\textwidth]{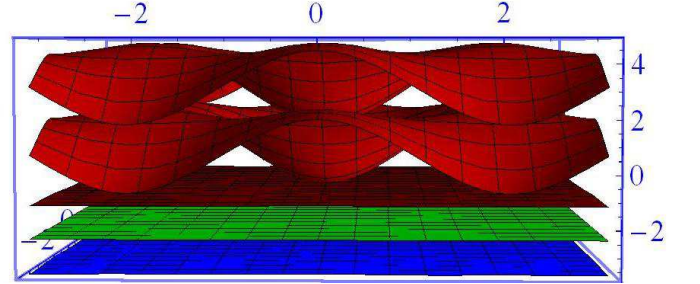}\caption{(Color online) The
electronic dispersion at the quarter filling ($d=0.5$) for the case with the
parameters $t^{\prime}=0.6t,$ $t^{\prime \prime}=-0.58t$ and $\phi_{ij}=0.4\pi
$, $U/t=5.0.$ The chemical potential $\mu_{eff}$ (the green plane) locates at
the middle of the energy gap. The ground state is a $Q=1$ topological
insulator with filled lowest topological flat band. }%
\end{figure}

\begin{figure}[ptb]
\includegraphics[width=0.5\textwidth]{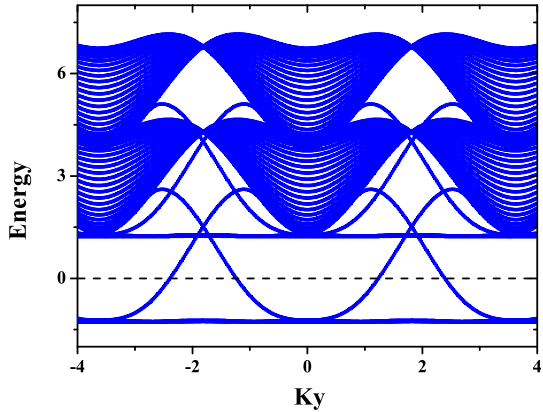}\caption{(Color online) The
Zigzag edge state along x-direction at $U/t=5.0$ with $t^{\prime}=0.6t$,
$t^{\prime \prime}=-0.58t$, $\phi_{ij}=0.4\pi$ for the quarter filling case
$d=0.5$. The black dash line is the position of chemical potential. }%
\end{figure}

\begin{figure}[ptb]
\includegraphics[width=0.5\textwidth]{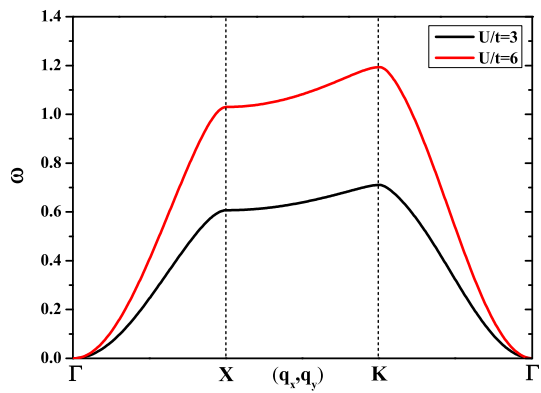}\caption{(Color online) The
dispersions of the spin wave of the FM order for the parameters are
$t^{\prime}=0.6t$, $t^{\prime \prime}=-0.58t$, $\phi_{ij}=0.4\pi$ with
$U/t=3.0$ and $U/t=6.0$ at $d=0.5$.}%
\end{figure}

Secondly we consider the quarter filling case: $d=0.5$. In the global phase
diagram in Fig.6, there exists an important line - the $d=0.5$ line. We found
that at the quarter filling the ground state is a $Q=1$\ topological insulator
induced by the FM order $M_{F}\neq0$. In Fig.6 we can see that there exist two
phases divided by a quantum phase transition at $U/t=0.1$ for $d=0.5$ : the
paramagnetic metallic state and the $Q=1$ topological insulator. In the weakly
interacting case $\left(  U/t<0.1\right)  $, the ground state is paramagnetic
metallic state with flat-band. Small interaction $\left(  U/t>0.1\right)  $
will drive the ground state to a $Q=1$ topological insulator with FM order.
From this result one may guess that in flat-band limit (the flatness ratio is
infinite), there exists only $Q=1$ topological insulator for $d=0.5$ and
quantum phase transition between the paramagnetic metallic state and the $Q=1$
topological insulator occurs at $U/t=0^{+}$. This result is consistent to the
rigorous proof in Ref.\cite{ka}, in which it was proved that the ground state
of the Hubbard model with topological flat-bands is a ferromagnetic order when
the lowest flat band is half-filled (that corresponds to the quarter filling
case here).

In Fig.12 and Fig.13, we show the ferromagnetization $M_{F}$ and the energy
gap for the case of $d=0.5$. In $Q=1$ topological insulator, the
ferromagnetization suddenly jumps to $0.5$ and gets saturated. In Fig.14, we
plot the electronic dispersion of the case $d=0.5$ in $Q=1$ topological
insulator region. From Fig.14, we can see clearly that the chemical potential
$\mu_{_{eff}}$ located at the middle of the energy gap, and the lowest band is
filled, while other bands are empty. We also calculate the edge states of
$Q=1$ topological insulator. Fig.15 illustrates the edge states of the
topological state. This result confirms the topological property of this $Q=1$
topological insulator. In this phase, we also go beyond the mean field theory
to study the collective spin fluctuations by RPA method (See Fig.16). In
Fig.16, we can see that although the dispersion of electrons is flat, the
collective spin modes are dispersive due the dispersive of the Matsubara
Green's function (See the detailed calculation in Appendix). However, at low
energy limit, the situation of FM order is much different from that of the
half filling case of AF-SDW order parameter : around the $\Gamma$ point, the
dispersion shows quadratic behavior instead of a linear one as $\omega
(k)\propto q^{2}$.

\section{Other cases}

\begin{figure}[ptb]
\includegraphics[width=0.5\textwidth]{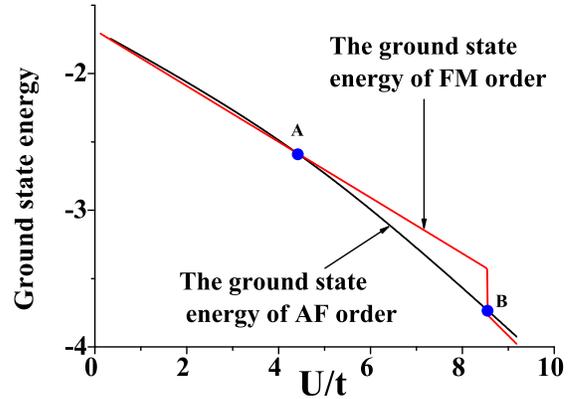}\caption{(Color online) The
ground state energies of AF order and FM order per site for the case of
$d=0.3$ at $t^{\prime}=0.6t,$ $t^{\prime \prime}=-0.58t$ and $\phi_{ij}%
=0.4\pi.$ The red line is the energy of FM order and the black line is the
energy of AF order. The blue spots $A$ and $B$ denote the quantum phase
transitions between AF order and FM order.}%
\end{figure}

\begin{figure}[ptb]
\includegraphics[width=0.5\textwidth]{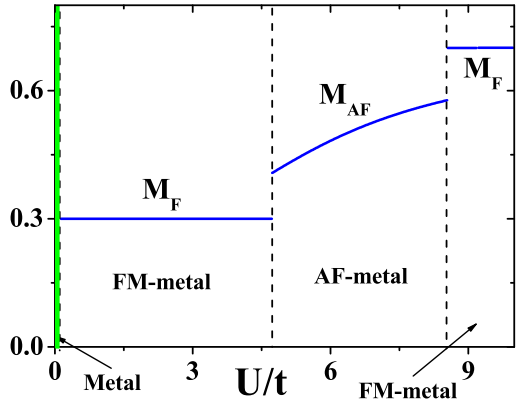}\caption{((Color online) The
magnetization ($M_{F}$ or $M_{AF}$) for the doping $d=0.3$ at $t^{\prime
}=0.6t,$ $t^{\prime \prime}=-0.58t$ and $\phi_{ij}=0.4\pi.$ There have three
phases: paramagnetic Metal (the Green region), FM-metal and AF-metal.}%
\end{figure}

\begin{figure}[ptb]
\includegraphics[width=0.62\textwidth]{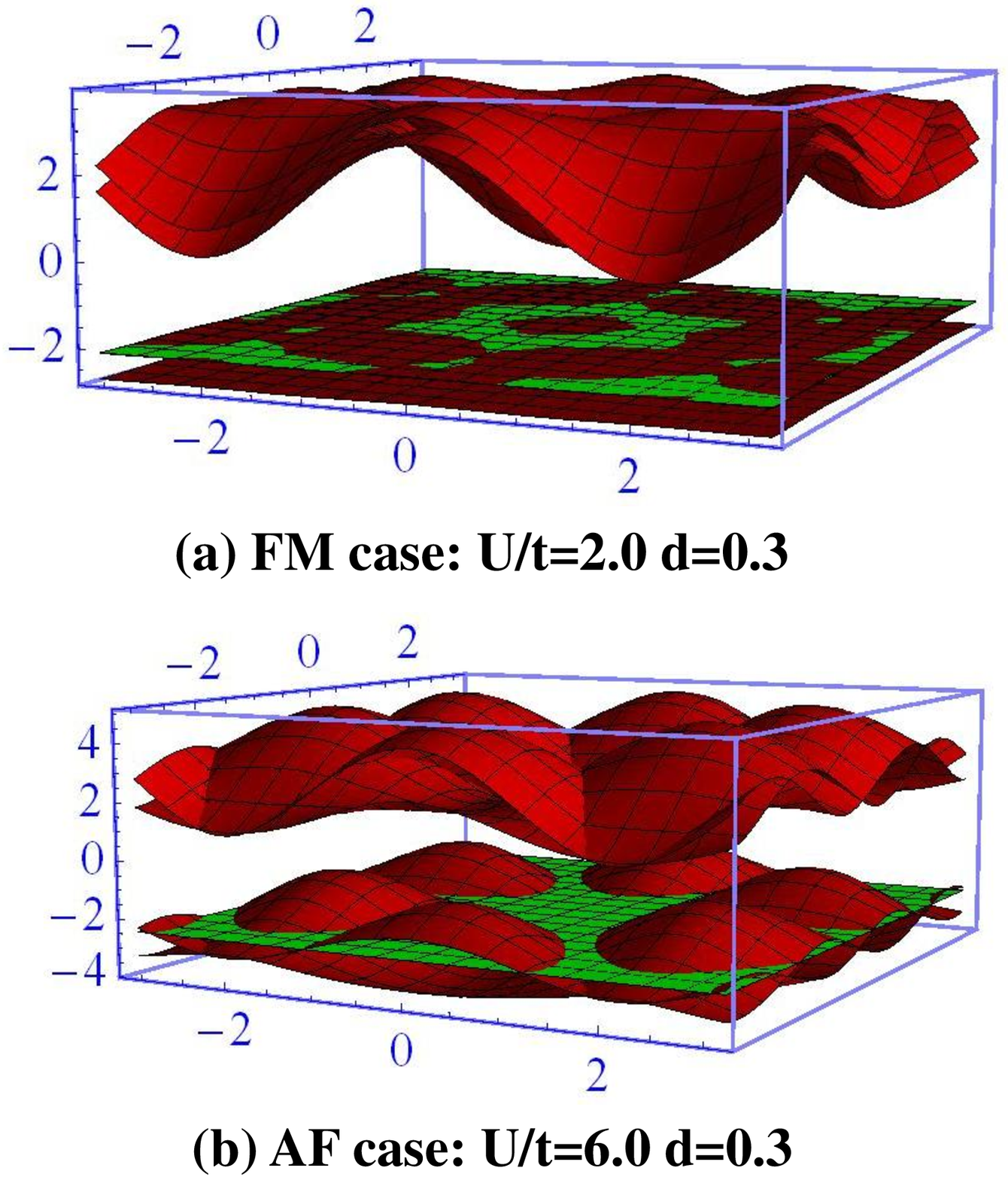}\caption{((Color online) The
electronic dispersion of the FM order and AF order at the doping $d=0.3$ for
$t^{\prime}=0.6t,$ $t^{\prime \prime}=-0.58t$ and $\phi_{ij}=0.4\pi.$ (a) is
the FM case and (b) is the AF case, we can see clearly that $\mu_{eff}$ (the
green plane) locate in the spectrum.}%
\end{figure}Thirdly, we consider other cases, $d\neq0,$ $d\neq0.5$. Here we
take $d=0.3$ as an example.

From Fig.6, we can see that in the weak interaction case $\left(
U/t<0.13\right)  $, the ground state is normal metallic state with flat-band.
Small interaction $\left(  U/t>0.13\right)  $ will drive the ground state to a
FM metallic state. The phase transition between normal metal with $M_{F}=0$
and FM-metal with $M_{F}\neq0$ is first order. The FM\ order parameter jumps
up at $U/t=0.13$. When we increase interaction strength, the ground state
energy for every site of the FM order is larger than AF order, the ground
state becomes the AF order crossing a first order phase transition at
$U/t=4.73$. An interesting property is that when we further increase
interaction strength the ground state energy for every site of the FM order is
lower than AF order again, the ground state re-enter the FM order crossing
another first order phase transition at $U/t=8.55$. In Fig.17, we give the
ground state energy comparation between AF order and FM order per site at zero
temperature when $d=0.3,$\ and we can see that there indeed have two
transitions points $A$ and $B$. In Fig.18 we show the magnetic order
parameters. In Fig.19 we also plot the electronic dispersion at the same
doping to compare the different dispersion in AF order and that in FM order.
Correspondingly, the electronic band structure changes when phase transitions occur.

\section{Conclusion and discussion}

In this paper, we investigated the magnetic properties (including
ferromagnetism and antiferromagnetism) of a correlated topological insulator
with flat band - the interacting spinful Haldane model by the mean field
approach. We found that the AF order is more stable than FM order at (or near)
half-filling due to energy gaining from super-exchange effect; while away from
half-filling, FM order is more stable than AF order due to the flat band. In
particular, at doping $d=0.5,$ the system becomes a $Q=1$ topological
insulator which is induced by the FM order. In addition, we found that in FM
order the degenerate flat bands split without changing the flatness ratio,
while in AF-SDW order, the electronic band structure changes totally and there
doesn't exist flat-band at all. We also used the random-phase-approximation to
calculate the dispersion of spin collective modes and found that in both FM
order and AF\ order, the spin collective modes are dispersive. In AF order,
the spin waves show linear dispersion while in FM order, the spin waves show
quadratic dispersion.

\begin{acknowledgments}
This work is supported by NFSC Grant No. 11174035, National Basic Research
Program of China (973 Program) under the grant No. 2011CB921803, 2012CB921704.
\end{acknowledgments}

\section{Appendix: RPA calculations for spin collective modes}

In this appendix, we use random-phase-approximation (RPA) to calculate the
spin dispersion at doping $d=0$ and $d=0.5$. The spin excitations are obtained
from the poles of the transverse spin susceptibility tensor, $\chi$, which is
defined, in Matsubara form, as
\[
\chi_{ij}^{+-}\left(  q,i\omega_{n}\right)  =\int_{0}^{\beta}d\tau
e^{i\omega_{n}\tau}\chi_{ij}^{+-}\left(  q,\tau \right)  ,
\]
where
\[
\chi_{ij}^{+-}\left(  q,\tau \right)  =\frac{1}{N_{s}}\left \langle T\left[
\hat{S}_{i,q}^{+}\left(  \tau \right)  ,\hat{S}_{j,-q}^{-}\left(  0\right)
\right]  \right \rangle
\]
and $i,j=a,b$ label the two sublattices and $S_{q}^{+}=\sum_{k}%
c_{_{k+q,\uparrow}}^{\dagger}c_{_{k,\downarrow}},\ S_{-q}^{-}=\sum
_{k}c_{_{k-q,\downarrow}}^{\dagger}c_{_{k,\uparrow}}$ denote the spin
operators for each sublattice. Using the Wick's theorem, we can derive the
leading order spin suspectibility in the AF phase and FM phase as%
\begin{align}
\chi_{ij}^{+-\left(  0\right)  }\left(  q,i\omega_{n}\right)   &  =-\frac
{1}{\beta N_{s}}\sum_{k,\omega_{m}}\emph{G}_{\uparrow}^{ji}\left(
k,i\omega_{m}\right) \nonumber \\
&  \times \emph{G}_{\downarrow}^{ij}\left(  k-q,i\omega_{n}+i\omega_{m}\right)
.
\end{align}

In AF phase, the Matsubara Green's functions are :%
\begin{align}
\emph{G}_{\sigma}^{aa}\left(  i\omega,k\right)   &  =\sum_{j=\pm}\frac
{\frac{1}{2}\left(  1+\frac{t^{\prime}D+\sigma \Delta_{A}}{\xi_{\sigma
,j}\left(  k\right)  }\right)  }{i\omega-E_{\sigma,j}\left(  k\right)  },\\
\emph{G}_{\sigma}^{bb}\left(  i\omega,k\right)   &  =\sum_{j=\pm}\frac
{\frac{1}{2}\left(  1-\frac{t^{\prime}D+\sigma \Delta_{A}}{\xi_{\sigma
,j}\left(  k\right)  }\right)  }{i\omega-E_{\sigma,j}\left(  k\right)  },\\
\emph{G}_{\sigma}^{ab}\left(  i\omega,k\right)   &  =\sum_{j=\pm}\frac
{\frac{\xi_{k}}{2\xi_{\sigma,j}\left(  k\right)  }}{i\omega-E_{\sigma
,j}\left(  k\right)  },\\
\emph{G}_{\sigma}^{ba}\left(  i\omega,k\right)   &  =\sum_{j=\pm}\frac
{\frac{\xi_{k}^{\ast}}{2\xi_{\sigma,j}\left(  k\right)  }}{i\omega
-E_{\sigma,j}\left(  k\right)  }%
\end{align}
where%
\begin{align*}
E_{\sigma,\pm}\left(  k\right)   &  =\pm \sqrt{\left(  tA+t^{\prime \prime
}E\right)  ^{2}+\left(  tB+t^{\prime \prime}F\right)  ^{2}+\left(  t^{\prime
}D+\sigma \Delta_{A}\right)  ^{2}}\\
&  -t^{\prime}C-\mu_{eff}%
\end{align*}%
\[
\xi_{\sigma,\pm}\left(  k\right)  =\pm \sqrt{\left(  tA+t^{\prime \prime
}E\right)  ^{2}+\left(  tB+t^{\prime \prime}F\right)  ^{2}+\left(  t^{\prime
}D+\sigma \Delta_{A}\right)  ^{2}}%
\]%
\[
\xi_{k}=-\left(  tA+t^{\prime \prime}E\right)  -i\left(  tB+t^{\prime \prime
}F\right)
\]

We may use the RPA approach to get the spin susceptibility tensor in AF order
from the Dyson equation
\begin{align}
\chi &  =\chi^{0}+U\chi^{0}\chi \  \nonumber \\
&  \Rightarrow \  \chi=\Big[\hat{I}-U\chi^{0}\Big]^{-1}\chi^{0}%
\end{align}
where $\hat{I}$ denotes the $2\times2$ identity matrix. The poles of the spin
susceptibility tensor, corresponding to the spin excitations, are then
obtained from the condition :
\begin{equation}
\mathrm{Det}\Big[\hat{I}-U\chi^{0}\Big]=0. \label{det}%
\end{equation}
We note that the tensorial nature of the spin susceptibility is a consequence
of two sublattices here.

The dispersion of spin collective modes with frequency $\omega$ and momentum
${q}$ is determined by above condition (\ref{det}) after performing the
analytic continuation $i\omega \rightarrow \omega+i0^{+}$. Finally the spin wave
dispersion determinant is given by
\begin{align}
D_{+-}(q,\omega)  &  =1-U\chi_{aa}^{+-\left(  0\right)  }-U\chi_{bb}%
^{+-\left(  0\right)  }\nonumber \\
&  +U^{2}\chi_{aa}^{+-\left(  0\right)  }\chi_{bb}^{+-\left(  0\right)
}-U^{2}\chi_{ab}^{+-\left(  0\right)  }\chi_{ba}^{+-\left(  0\right)
}\nonumber \\
&  =0,
\end{align}
where the elements of the spin susceptibility tensor $\chi^{+-\left(
0\right)  }$ are :\begin{widetext}
\begin{align}
\chi_{aa}^{+-\left(  0\right)  }\left(  q,\omega_{n}\right)   &  =-\frac
{1}{N_{s}}\sum_{k}\Big[\left(  \frac{\left(  1-\frac{t^{\prime}D\left(
k\right)  +\Delta_{A}}{\xi_{\uparrow}\left(  k\right)  }\right)  \left(
1+\frac{t^{\prime}D\left(  k-q\right)  -\Delta_{A}}{\xi_{\downarrow}\left(
k-q\right)  }\right)  }{4\left(  \omega_{n}-E_{\downarrow,+}\left(
k-q\right)  +E_{\uparrow,-}\left(  k\right)  \right)  }+\frac{\left(
1-\frac{t^{\prime}D\left(  k\right)  +\Delta_{A}}{\xi_{\uparrow}\left(
k\right)  }\right)  \left(  1-\frac{t^{\prime}D\left(  k-q\right)  -\Delta
_{A}}{\xi_{\downarrow}\left(  k-q\right)  }\right)  }{4\left(  \omega
_{n}-E_{\downarrow,-}\left(  k-q\right)  +E_{\uparrow,-}\left(  k\right)
\right)  }\right)  \theta \left(  -E_{\uparrow,-}\left(  k\right)  \right)
\nonumber \\
&  -\left(  \frac{\left(  1+\frac{t^{\prime}D\left(  k\right)  +\Delta_{A}%
}{\xi_{\uparrow}\left(  k\right)  }\right)  \left(  1-\frac{t^{\prime}D\left(
k-q\right)  -\Delta_{A}}{\xi_{\downarrow}\left(  k-q\right)  }\right)
}{4\left(  \omega_{n}-E_{\downarrow,-}\left(  k-q\right)  +E_{\uparrow
,+}\left(  k\right)  \right)  }+\frac{\left(  1-\frac{t^{\prime}D\left(
k\right)  +\Delta_{A}}{\xi_{\uparrow}\left(  k\right)  }\right)  \left(
1-\frac{t^{\prime}D\left(  k-q\right)  -\Delta_{A}}{\xi_{\downarrow}\left(
k-q\right)  }\right)  }{4\left(  \omega_{n}-E_{\downarrow,-}\left(
k-q\right)  +E_{\uparrow,-}\left(  k\right)  \right)  }\right)  \theta \left(
-E_{\downarrow,-}\left(  k-q\right)  \right)  \Big]
\end{align}
\end{widetext}\begin{widetext}
\begin{align}
\chi_{bb}^{+-\left(  0\right)  }\left(  q,\omega_{n}\right)   &  =-\frac
{1}{N_{s}}\sum_{k}\Big[\left(  \frac{\left(  1+\frac{t^{\prime}D\left(
k\right)  +\Delta_{A}}{\xi_{\uparrow}\left(  k\right)  }\right)  \left(
1-\frac{t^{\prime}D\left(  k-q\right)  -\Delta_{A}}{\xi_{\downarrow}\left(
k-q\right)  }\right)  }{4\left(  \omega_{n}-E_{\downarrow,+}\left(
k-q\right)  +E_{\uparrow,-}\left(  k\right)  \right)  }+\frac{\left(
1+\frac{t^{\prime}D\left(  k\right)  +\Delta_{A}}{\xi_{\uparrow}\left(
k\right)  }\right)  \left(  1+\frac{t^{\prime}D\left(  k-q\right)  -\Delta
_{A}}{\xi_{\downarrow}\left(  k-q\right)  }\right)  }{4\left(  \omega
_{n}-E_{\downarrow,-}\left(  k-q\right)  +E_{\uparrow,-}\left(  k\right)
\right)  }\right)  \theta \left(  -E_{\uparrow,-}\left(  k\right)  \right)
\nonumber \\
&  -\left(  \frac{\left(  1-\frac{t^{\prime}D\left(  k\right)  +\Delta_{A}%
}{\xi_{\uparrow}\left(  k\right)  }\right)  \left(  1+\frac{t^{\prime}D\left(
k-q\right)  -\Delta_{A}}{\xi_{\downarrow}\left(  k-q\right)  }\right)
}{4\left(  \omega_{n}-E_{\downarrow,-}\left(  k-q\right)  +E_{\uparrow
,+}\left(  k\right)  \right)  }+\frac{\left(  1+\frac{t^{\prime}D\left(
k\right)  +\Delta_{A}}{\xi_{\uparrow}\left(  k\right)  }\right)  \left(
1+\frac{t^{\prime}D\left(  k-q\right)  -\Delta_{A}}{\xi_{\downarrow}\left(
k-q\right)  }\right)  }{4\left(  \omega_{n}-E_{\downarrow,-}\left(
k-q\right)  +E_{\uparrow,-}\left(  k\right)  \right)  }\right)  \theta \left(
-E_{\downarrow,-}\left(  k-q\right)  \right)  \Big]
\end{align}
\end{widetext}
\begin{widetext}
\begin{align}
\chi_{ba}^{+-\left(  0\right)  }\left(  q,\omega_{n}\right)   &  =-\frac
{1}{N_{s}}\sum_{k}\frac{\xi_{k}\xi_{k-q}^{\ast}}{4\xi_{\uparrow}\left(
k\right)  \xi_{\downarrow}\left(  k-q\right)  }\Big[\left(  \frac{1}%
{\omega_{n}-E_{\downarrow,-}\left(  k-q\right)  +E_{\uparrow,-}\left(
k\right)  }-\frac{1}{\omega_{n}-E_{\downarrow,+}\left(  k-q\right)
+E_{\uparrow,-}\left(  k\right)  }\right)  \theta \left(  -E_{\uparrow
,-}\left(  k\right)  \right)  \nonumber \\
&  -\left(  \frac{1}{\omega_{n}-E_{\downarrow,-}\left(  k-q\right)
+E_{\uparrow,-}\left(  k\right)  }-\frac{1}{\omega_{n}-E_{\downarrow,-}\left(
k-q\right)  +E_{\uparrow,+}\left(  k\right)  }\right)  \theta \left(
-E_{\downarrow,-}\left(  k-q\right)  \right)  \Big]
\end{align}
\end{widetext}
\begin{widetext}
\begin{align}
\chi_{ab}^{+-\left(  0\right)  }\left(  q,\omega_{n}\right)   &  =-\frac
{1}{N_{s}}\sum_{k}\frac{\xi_{k}^{\ast}\xi_{k-q}}{4\xi_{\uparrow}\left(
k\right)  \xi_{\downarrow}\left(  k-q\right)  }\Big[\left(  \frac{1}%
{\omega_{n}-E_{\downarrow,-}\left(  k-q\right)  +E_{\uparrow,-}\left(
k\right)  }-\frac{1}{\omega_{n}-E_{\downarrow,+}\left(  k-q\right)
+E_{\uparrow,-}\left(  k\right)  }\right)  \theta \left(  -E_{\uparrow
,-}\left(  k\right)  \right)  \nonumber \\
&  -\left(  \frac{1}{\omega_{n}-E_{\downarrow,-}\left(  k-q\right)
+E_{\uparrow,-}\left(  k\right)  }-\frac{1}{\omega_{n}-E_{\downarrow,-}\left(
k-q\right)  +E_{\uparrow,+}\left(  k\right)  }\right)  \theta \left(
-E_{\downarrow,-}\left(  k-q\right)  \right)  \Big]
\end{align}
\end{widetext}

In FM order, we may use similar approach to calculate the spin suspectibility
and the dispersion of spin collective modes.

\end{document}